\documentclass[conference]{IEEEtran}
\IEEEoverridecommandlockouts
\usepackage[
maxnames=99,
  backend=biber,
  style=ieee,         
  citestyle=numeric-comp,
  sorting=nyt        
]{biblatex}
\usepackage{float}
\usepackage{amsmath,amssymb,amsfonts}
\usepackage{algorithmic}
\usepackage{graphicx}
\usepackage{textcomp}
\usepackage{xcolor}
\usepackage{multirow}
\usepackage{booktabs}
\usepackage{adjustbox}
\usepackage{graphicx}
\usepackage{subcaption}
\def\BibTeX{{\rm B\kern-.05em{\sc i\kern-.025em b}\kern-.08em
    T\kern-.1667em\lower.7ex\hbox{E}\kern-.125emX}}
\begin{document}

\title{Confidence Driven Classification of Application Types in the Presence of Background Network Traffic\thanks{This material is based upon work supported by the National Science Foundation under Grant No. IMR-2319511.}}
\author{%
  \begin{minipage}[t]{0.30\textwidth}\centering
    Eun Hun Choi\\
    Computer Science\\
    UNC Chapel Hill, United States\\
    paulchoi@cs.unc.edu
  \end{minipage}%
  \hfill
  \begin{minipage}[t]{0.30\textwidth}\centering
    Jasleen Kaur\\
    Computer Science\\
    UNC Chapel Hill, United States\\
    jasleen@cs.unc.edu
  \end{minipage}%
  \hfill
  \begin{minipage}[t]{0.30\textwidth}\centering
    Vladas Pipiras\\
    Statistics\\
    UNC Chapel Hill, United States\\
    pipiras@email.unc.edu
  \end{minipage}

  \\[2ex] 

  \begin{minipage}[t]{0.45\textwidth}\centering
    Nelson Gomes\\
    Statistics\\
    Universidade do Algarve, Faro, Portugal\\
    nantunes@ualg.pt
  \end{minipage}%
  \hfill
  \begin{minipage}[t]{0.45\textwidth}\centering
    Brendan Massey\\
    Computer Science\\
    UNC Chapel Hill, United States\\
    masseybr@cs.unc.edu
  \end{minipage}
}

\maketitle

\begin{abstract}
Accurately classifying the application types of network traffic using deep learning models has recently gained popularity. However, we find that these classifiers do not perform well on real-world traffic data due to the presence of non-application-specific generic background traffic originating from advertisements, analytics, shared APIs, and trackers. Unfortunately, state-of-the-art application classifiers overlook such traffic in curated datasets and only classify relevant application traffic. To address this issue, when we label and train using an additional class for background traffic, it leads to additional confusion between application and background traffic, as the latter is heterogeneous and encompasses {\em all} traffic that is not relevant to the application sessions. To avoid falsely classifying background traffic as one of the relevant application types, a reliable confidence measure is warranted, such that we can refrain from classifying uncertain samples. Therefore, we design a Gaussian Mixture Model-based classification framework that improves the indication of the deep learning classifier's confidence to allow more reliable classification.  
\end{abstract}


\section{Introduction}
Identifying the type of application associated with network traffic is important for understanding overall usage patterns, monitoring network performance, and ensuring equitable resource allocation for different applications. Early methods of application classification of network traffic included port number identification and deep packet inspection \cite{survey}. The increase in the use of dynamic port numbers as well as traffic encryption reduced the reliability as well as feasibility of these \cite{survey}. Instead, recent network traffic classification relies on training deep learning models based on structural traffic features, unencrypted traffic header fields, and/or encrypted payload bytes \cite{survey,malware,fs,flowpic,etbert,akbari}.

Deep learning classifiers have been found to perform very well on carefully curated popular benchmarks like ISCX VPN-nonVPN and UC-Davis QUIC \cite{iscx,quic}. However, when we collected a more recent traffic from contemporary application types for evaluation, state-of-the-art classifiers did not perform well. To identify the issue, we examined the DNS query/response preceding each traffic flow, and we found that many applications communicated with domain names that indicated shared APIs, advertisements, and analytics that are not necessarily unique to the application. The presence of such generic {\em background} traffic in the training/test data confuses state-of-the-art classifiers---such traffic is either not included or specifically filtered in the popular curated datasets, but will be encountered when a classifier is used in the real world. We then label all such background traffic flows as a separate background class in our training and test datasets. This improved the model's performance, but we still observed notable misclassifications between application and background traffic.

It is undesirable for background traffic to be labeled as an application traffic by a classifier, or for relevant application traffic to be incorrectly identified as background traffic. These misclassifications reduce both the performance and coverage needed for monitoring the usage, performance, and resource needs of different applications. In such scenarios, instead, it may be preferable that the deep learning model assign labels to traffic flow only when it can do so with high confidence. Therefore, in this paper, we develop a framework that can characterize the confidence of a model's classification decision in a manner that significantly improves the classification performance. Our main innovations in this paper are as follows:
\begin{enumerate}
\item We address the presence of background traffic by designing a classifier that can handle such traffic in the wild.
\item We investigate the trade-off between performance and coverage in traffic classification and subsequently design a classification framework that can help achieve different operating points in this trade-off based on the confidence threshold.  
\end{enumerate}
In the rest of the paper, we describe our dataset in Section II, present deep learning models in Section III, conduct baseline experiments in Section IV, consider uncertainty using softmax in Section V, introduce a Gaussian Mixture Model classification framework in Section VI, discuss related work in Section VII, and conclude in Section VIII.

\begin{table*}[htbp]
  \resizebox{\textwidth}{!}{%
    \begin{tabular}{|c|c|p{7.48cm}|c|c|}
      \hline
      & \textbf{App Types} & \textbf{Application Instances} & \textbf{Train Sessions} & \textbf{Test Sessions}\\
      \hline
      \multirow{4}{*}{\rotatebox[origin=c]{90}{Selenium \ \ \ }}
        & Web Browsing  & Google, Baidu, Wikipedia, Yandex, Naver, Docomo, Samsung, Weather, Microsoft, DuckDuckGo, eBay, ESPN, CNN, Etsy, AOL, USPS, Yahoo 
          & 314 & 104 \\
      \cline{2-5}
        & Social Media  & Quora, Facebook, Instagram, Twitter, Reddit, Pinterest 
          & 72 & 30\\
      \cline{2-5}
        & Video Streaming & YouTube, Twitch, Netflix 
          & 39 & 15\\
      \cline{2-5}
        & Email  & Outlook, Gmail 
          & 46 & 24\\
      \hline\hline
      \multirow{4}{*}{\rotatebox[origin=c]{90}{Manual}}
        & VoIP/Video Call & Skype, Facebook, Zoom 
          & 129 & 67\\
      \cline{2-5}
        & Chat  & WhatsApp, Messenger 
          & 74 & 30\\
      \cline{2-5}
        & Gaming  & FC24, Overwatch 
          & 65 & 19\\
      \cline{2-5}
        & Online Doc  & Google Docs, Microsoft Word 
          & 25 & 13\\
      \hline\hline
      \multicolumn{3}{|c|}{\textbf{Total Sessions}} 
        & \textbf{764} & \textbf{302}\\
      \hline
    \end{tabular}%
  }
  \vspace{0.01mm}
  \caption{Application Types and Number of Sessions in Our Dataset}
  \label{tab:apptypes}
  \vspace{-7mm}
\end{table*}

\section{Data Preparation Methodology}
\label{sec:data-collection}
We first summarize how existing traffic datasets used in classification studies are curated. We then introduce the methodology for collecting our own dataset.
\subsection{Popular Datasets}
ISCX VPN-nonVPN is used in numerous studies and contains traffic captured from 7 application types: web browsing, email, chat, streaming, file transfer, VoIP, and TraP2P \cite{iscx}. For data collection, the authors manually performed different tasks in the respective applications and saved the captured traffic in the standard PCAP format. To make sure that each PCAP file solely contains traffic generated from the specific application software, they closed unnecessary applications. 
The UC-Davis QUIC dataset has traffic collected from 5 Google services: Drive, Youtube, Docs, Search, and Music \cite{quic}. To allow large-scale data collection, automated scripts using Selenium and Autoit were used to emulate human behaviors while capturing the traffic. Additionally, several samples of traffic generated from real human interactions were included. All non-QUIC traffic was filtered out from the dataset. While popular, ISCX and QUIC datasets were collected in 2016 and 2018, respectively, and are not sufficiently recent.

TLS encrypted traffic datasets have also been used for application classification. A proprietary dataset (Orange '20) is used in \cite{akbari}, where 119,565 out of 343,228 TLS traffic flows are labeled using Server Name Indication (SNI). TlS 1.3 dataset is used in \cite{etbert} to specifically identify and label TLS traffic for 120 different applications based on SNI. 
\subsection{Our Data Collection}
In order to collect a more recent dataset, we tapped into modern Internet traffic. We identified popular application types using the 2024 Global Internet Phenomena Report \cite{sandvine2024}, according to which video streaming (48\%), social media (10\%), and device gaming (9\%) had the highest downstream traffic volume in the Americas region, in addition to television. Additionally, we included several popular services that represent diverse user behaviors: general web browsing, email, VoIP, chat, and online document editing. 

To make our data collection scalable, we automated web browsing, social media, video, and email using Selenium \cite{selenium}. For web browsing, social media, and video, we selected 26 different websites from the top-50 most visited websites \cite{similarweb}, where actions could be automated with relative ease using Selenium (see Table \ref{tab:apptypes}). Our script emulated human behaviors using the following actions (with a hard coded gap ranging from 1-10 seconds between actions): 
\begin{itemize}

\item {\em Web browsing}: We visited different pages linked from the landing page of the website. If the website contained a search engine, we also searched for a randomly selected popular keyword from those listed in \cite{semrush}.

\item {\em Social media}: We created accounts for our research and scrolled through posts consisting of videos, images, and texts; we also put a reaction on randomly-selected posts. 

\item {\em Video streaming}: For YouTube, we first randomly picked from a list of most viewed videos and played it for 3-5 minutes \cite{popular}. For all three video streaming websites, we clicked on a random video appearing on the landing page and played it for 3-5 minutes. 

\item {\em Email}: We logged into the account and sent an email to a randomly-selected recipient. For Outlook, we also read the most recently-received email. 
\end{itemize}
We ran data collection using several application ``sessions,'' visiting only one website in each, and captured all traffic flows initiated. 

The remaining four application types are difficult to automate. We manually ran sessions as follows and captured the traffic generated in each session:
\begin{itemize}
\item 
{\em Chat \& VoIP:} We used two separate devices to initiate interactive sessions. 
\item 
{\em Online doc:} We made edits to an online document. 
\item 
{\em Gaming:} We played games on PlayStation 5.

\end{itemize}
Table \ref{tab:apptypes} summarizes the applications/websites we used for each application type and the 1066 application sessions we emulated and collected for our dataset. We split the application sessions into train and test datasets, which ensures that traffic in these was collected at different times with no overlap.

\subsection{Traffic Capture}
We capture {\em all} traffic initiated in each application session, including TCP and UDP transfers, and DNS.
For most application types, we began the traffic capture just before the application launch, so we only keep flows with a preceding DNS response (we disregard lingering flows from before). For VoIP and gaming, we began capturing traffic just after the session began, so we keep all traffic.\footnote{We ignore mDNS traffic (port number 5353) from the VoIP sessions.} Additionally, to mirror a realistic flow tracker, which marks a TCP session as complete once FIN flags have been exchanged in both directions, we disregard any TCP packets seen after both FINs have been observed. We also only study flows with at least 40 packets.
\section{Deep Learning Classifiers: State of the Art}
Various feature representations have been proposed for deep learning models. One of the earliest approaches simply fed the first $n$ raw bytes of each flow directly into a model for malware detection \cite{malware}. Building on this idea, another study \cite{discriminative} converted the first 512 bytes (beginning at the transport layer header) across 32 packets per flow into an image‐based representation for classification. As an alternative to feeding raw bytes of a flow directly into the model, ET-BERT \cite{etbert} segmented traffic into “BURST,” which is a sequence of consecutive packets traveling in the same direction. Each BURST’s raw datagram bytes were split into two-byte bigrams, and the bigrams were converted to tokens for model input \cite{etbert}. The model learned the contextual relationships among these tokens through self-supervised pre-training and was subsequently fine-tuned for different classification tasks \cite{etbert}. However, one prior work has found that classifiers trained on raw packet payload bytes in encrypted traffic may overfit on unencrypted handshake fields. In their evaluation, masking SNI alone caused accuracy of the classifier to fall to 38.19\%, and masking all handshake fields drove it down to 27.52\% \cite{snimask}. Supporting this concern, a different study \cite{etbertcritique} demonstrated that ET-BERT may overfit on specific packet-level information, ranging from SNI to session-specific header fields, when the dataset is not carefully curated. These findings highlight the importance of selecting robust features to ensure the model’s ability to generalize to traffic, where certain types
of packet-level information may be unavailable.

In contrast, other studies employed time series features
that are not dependent on specific packet bytes, such
as sequences of packet sizes, arrival times, and directions for
traffic classification \cite{flowpic,fs,akbari}. Since these features are universally available in TCP and UDP traffic\footnote{In this paper, we group UDP packets sharing the same 5-tuple and refer to them collectively as a UDP flow.} and help prevent overfitting on packet level bytes \cite{akbari}, we experimentally evaluate classifiers that use time series features. We use Flow Sequence Network (FS-Net) \cite{fs} and our implementation of a Bidirectional LSTM (BiLSTM) \cite{bilstm}, as FS-Net and various forms of LSTM have been used for evaluations in prior studies \cite{akbari,ode,rosetta,fs,etbert}. Each model is trained with cross-entropy loss, using a distinct combination of time series features and data balancing techniques. Although FS-Net overall achieves better performance, our BiLSTM yields comparable results, demonstrating that performance trends persist across different architectures, feature sets, and data balancing methods. Hence, we include both models in our evaluation. The architecture of our BiLSTM is as follows:
\begingroup
  \sloppy 
  \begin{itemize}
    \item \textbf{Input Layer}: Shape~$40\times2$, Data type~\texttt{float32}, Activation~none.
    \item \textbf{Bidirectional LSTM}: 128 units, \\ \texttt{return\_sequences=True}, Activation~none.
    \item \textbf{Layer Normalization}.
    \item \textbf{Bidirectional LSTM}: 64 units, \\ \texttt{return\_sequences=False}, Activation none.
    \item \textbf{Layer Normalization}.
    \item \textbf{Dropout}: Rate~0.1.
    \item \textbf{Dense}: 64~units, Activation~GELU.
    \item \textbf{Dense (Output)}: Activation~Softmax.
  \end{itemize}
\endgroup
We use the time series features defined in \cite{akbari} to train the BiLSTM. Each flow is represented by a 40 × 2 feature vector corresponding to the first 40 packets, where at each index, the two features are: packet arrival time relative to the flow start and packet size. To encode directionality, we negate the packet size of the server packets. Since our training dataset has unequal number of flows for each class, we use data augmentation to balance the number of flows, since appropriate data augmentation improves training \cite{augment}. We adapt the translation technique that works well with time series features \cite{augment} and derive an augmented feature vector from a sampled original one as follows: we randomly select a start index of a subsequence and shift it either to the left or to the right by \textit{n} steps. For left shifts, if the original traffic flow had more than 40 packets, we use the subsequent packets to fill in the new trailing empty \textit{n} rightmost slots in the feature vector. If there are not enough packets to fill the right most slots, we use zero padding to fill the remaining slots. For right shifts, we use the packet arrival time and packet size of the first index of the subsequence to fill in the vacated \textit{n} slots. 

For FS-Net, we use the original input of the 256 packet size sequence for our experiments. Also, we add a softmax layer at the end of the model so that it outputs softmax probabilities instead of logits (since we later use softmax probabilities for analysis). For consistency with the experiments using the BiLSTM, we use the same set of traffic flows and zero pad flows shorter than 256 packets. To balance the training dataset without modifying the original experimental setup through data augmentation, we oversample the minority classes so that each class has an equal number of traffic flows.
\section{Baseline Classification Evaluation}
In this section, we present baseline results using the labeling schemes from existing curated datasets, and we compare them against our own comprehensive labeling method.
\subsection{Classification Performance Metrics}
For evaluation, we use three different metrics for performance: macro average F1 score, average accuracy, and weighted F1 score.\footnote{Macro Average F1 = $\frac{1}{C} \sum_{c=1}^{C} F_1(c)$, where $C$ is the number of classes and $F_1(c)$ is the F1 score for class $c$.

\noindent Average Accuracy =  $\frac{1}{K} \sum_{i=1}^{K} \mathbb{I}(\hat{y}_i = y_i)$, where $K$ is the total number of samples, $\hat{y}_i$ is the predicted label, $y_i$ is the actual label, and the indicator function $\mathbb{I}(\hat{y}_i = y_i)$ is equal to 1 when $\hat{y}_i = y_i$ and 0 otherwise.

\noindent Weighted Average F1 = $\frac{\sum_{c=1}^{C} n_c \cdot F_1(c)}{\sum_{c=1}^{C} n_c}$, where $C$ is the number of classes, $n_c$ is the number of samples in class $c$, and $F_1(c)$ is the F1 score for class $c$.}
We include macro average F1 score because it accounts for the class imbalance in our test dataset by weighing the F1 score of each class equally. Additionally, we present an in‑depth performance analysis using a confusion matrix of the model that performed better for the corresponding experiment.   
\subsection{Application Session-Based Labeling Scheme}
For our first baseline experiment, we labeled the traffic strictly based on the application session it originated from, since the data collection was performed separately for each application session (as done in the ISCX and QUIC datasets). Table \ref{tab:session-based} shows the performance of the model.
\begin{table}[H]
  \centering
  \scalebox{1.2}{%
  \begin{tabular}{lccc}
    \hline
    \textbf{Model} & \textbf{Macro F1} & \textbf{Accuracy} & \textbf{Weighted F1} \\ 
    \hline
    BiLSTM              & 0.72              & 0.79              & 0.81                 \\
    \hline
    FS-Net &0.75 &0.78 &0.81\\
    \hline
  \end{tabular}
  }
  \caption{Performance: Session Based Labels}
  \label{tab:session-based}
\end{table}

\begin{figure}[H]
  \centering
  \centering
  \includegraphics[width=0.9\linewidth]{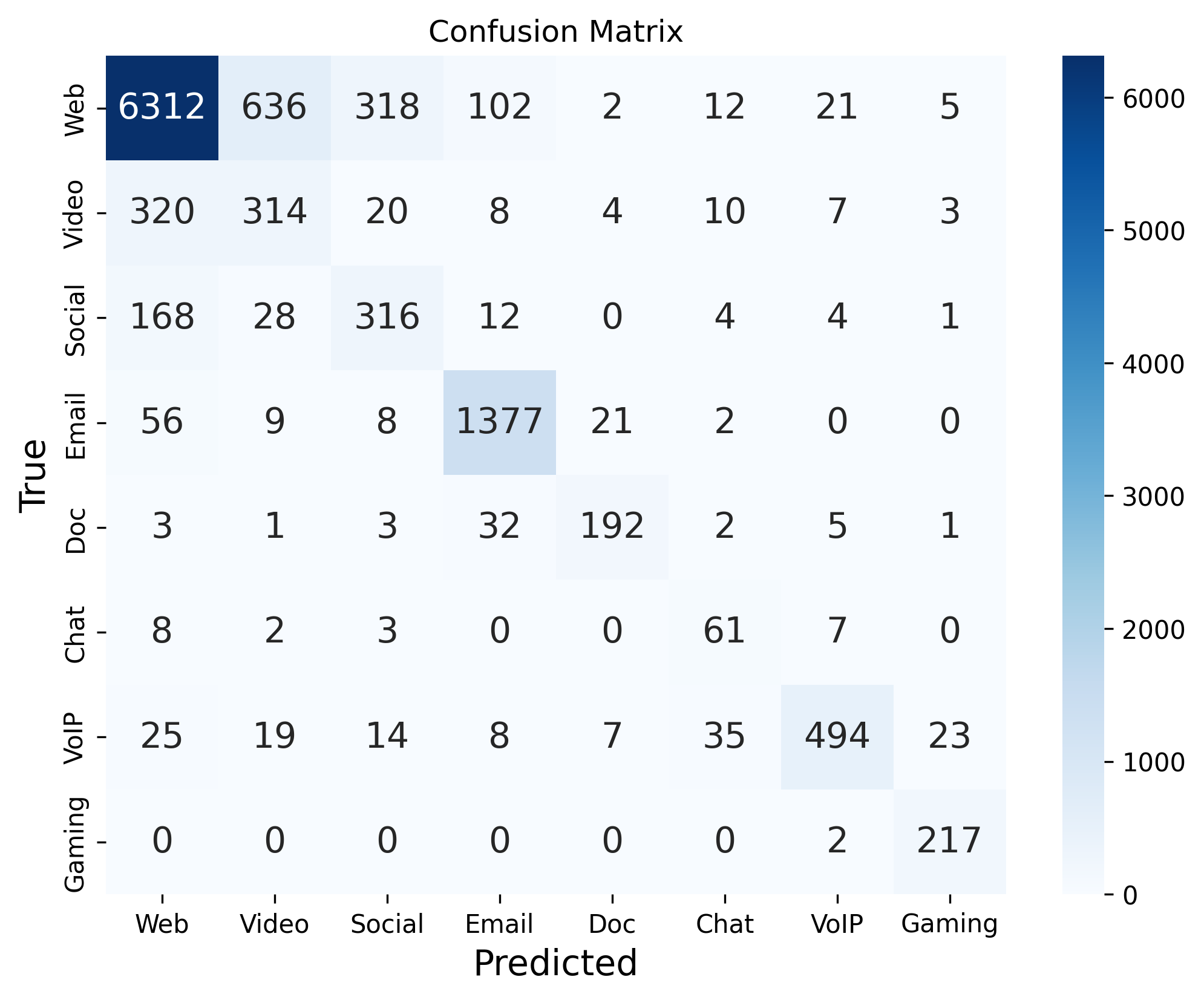}
  \caption{Session-Based Labeled Experiment (FS-Net)}
  \label{fig:session}
\end{figure}

Fig. \ref{fig:session} shows noticeable misclassifications in web browsing, video streaming, social media, and email traffic flows. The confusion among the four classes causes lower performance across the three metrics, with macro average F1 score particularly lower for both BiLSTM and FS-Net as seen in Table \ref{fig:session}. To further investigate the problem, we inspect the domain names from the DNS response for each traffic flow. We find that among applications that use the web browser, there is significant amount of traffic that is not necessarily unique to the application, such as shared APIs, analytics, advertisements, and trackers. Such ``background" traffic is not included in the curated datasets used in prior studies, and its presence may have led to the poor results we observe. 
\subsection{Domain Name-Based Labeling Scheme}
To confirm our suspicion, we create a curated version of our dataset using the methodology below. We adapt a simple flow-based labeling similar to SNI-based labeling, where if the substring of the domain name explicitly indicates the application type, we label it as corresponding application. For VoIP, we only keep the traffic flow responsible for two-way communication between the local devices. For gaming, we keep all traffic flows since all traffic come from the gaming console device. Also, we found that relevant traffic from same Azure domain appears across Microsoft services in different application types. To avoid confusion in labeling, we also introduce a new class, ``Azure” that groups traffic with same domain name from web browsing, email, and online document editing. Similar to other datasets, we filter rest of the traffic that is not relevant to the application type. Table \ref{tab:domain-based} shows the performance of the model. 
\begin{table}[H]
  \centering
  \scalebox{1.2}{%
  \begin{tabular}{lccc}
    \hline
     \textbf{Model} & \textbf{Macro F1} & \textbf{Accuracy} & \textbf{Weighted F1} \\
    \hline
     BiLSTM &0.92     & 0.95     & 0.95         \\
     \hline
     FS-Net &0.93 &0.96 &0.96\\
    \hline
  \end{tabular}
  }
    \caption{Performance: Domain Name-Based Labels}
  \label{tab:domain-based}
\end{table}
\begin{figure}[H]
  \centering
  \centering
  \includegraphics[width=0.9\linewidth]{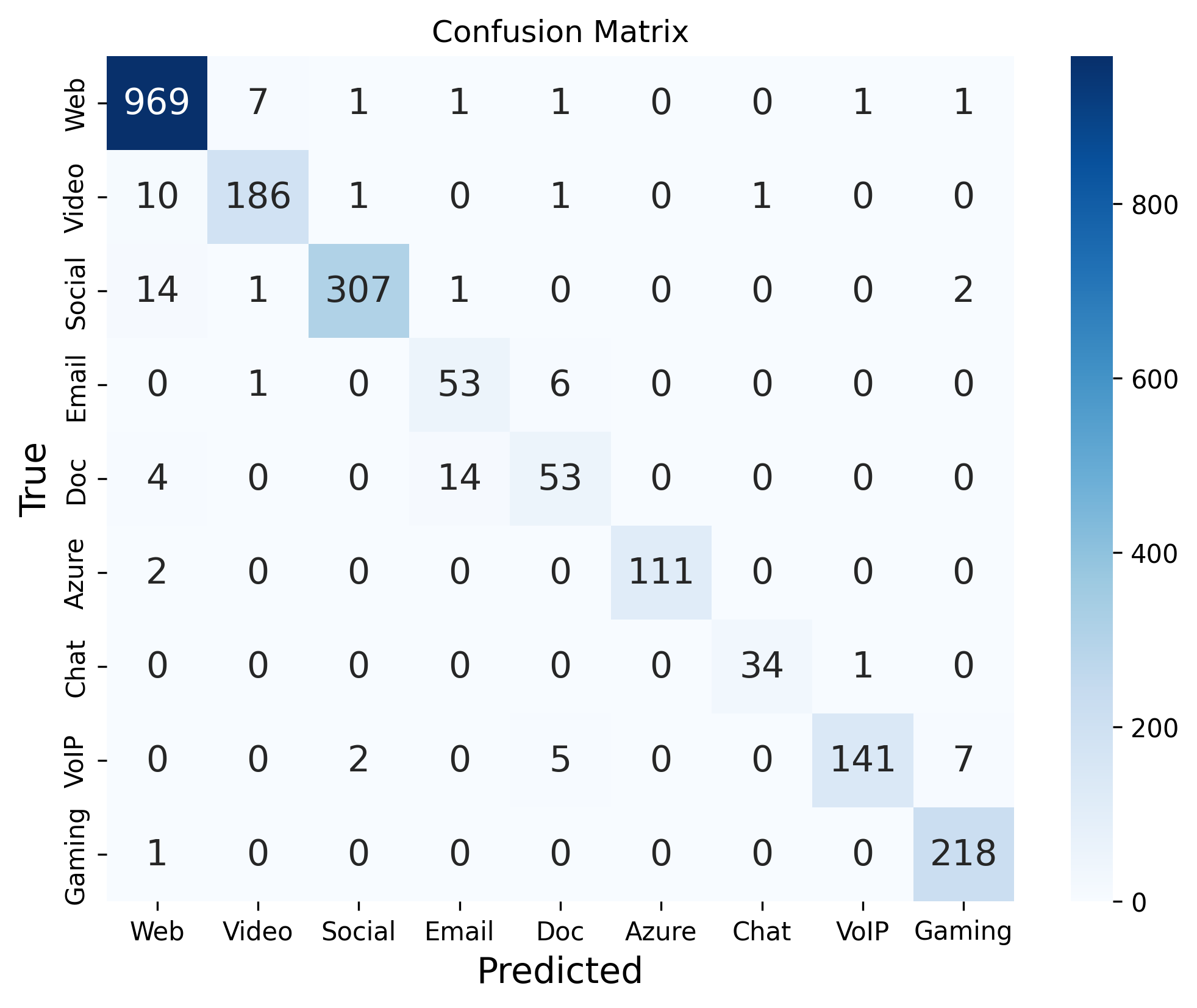}
  \caption{Domain Name-Based Labeled Experiment (FS-Net)}
  \label{fig:relevant}
\end{figure}

We observe better performance in the models with 0.93 and 0.92 macro average F1 score in FS-Net and BiLSTM, respectively, as seen in Table \ref{tab:domain-based}. This confirms our suspicion that the background traffic from different applications causes the performance degradation. However, this model does not handle the presence of background traffic. If background traffic is encountered in the wild, it will be misclassified as one of the application types.
\color{black}
\subsection{Comprehensive Domain Name-Based Labeling Scheme}
To accommodate the presence of background traffic without adversely affecting classification performance for application-specific traffic, we introduce a new class for the former. If no substring in the domain name indicates its relevance to the corresponding application session, we label it as background. If we separately label the background traffic and include it as an additional class to train the model, we get the following result in Table \ref{tab:comprehensive}.     
\begin{table}[h]
  \centering
  \scalebox{1.2}{%
  \begin{tabular}{lccc}
    \hline
     \textbf{Model} & \textbf{Macro F1} & \textbf{Accuracy} & \textbf{Weighted F1} \\
    \hline
     BiLSTM &0.84     & 0.94     & 0.94         \\
     \hline
    FS-Net &0.87 & 0.95 & 0.95\\
    \hline
  \end{tabular}
  }
    \caption{Performance: Comprehensive Labels}
  \label{tab:comprehensive}
\end{table}

\begin{figure}[h]
  \centering
  \centering
  \includegraphics[width=0.9\linewidth]{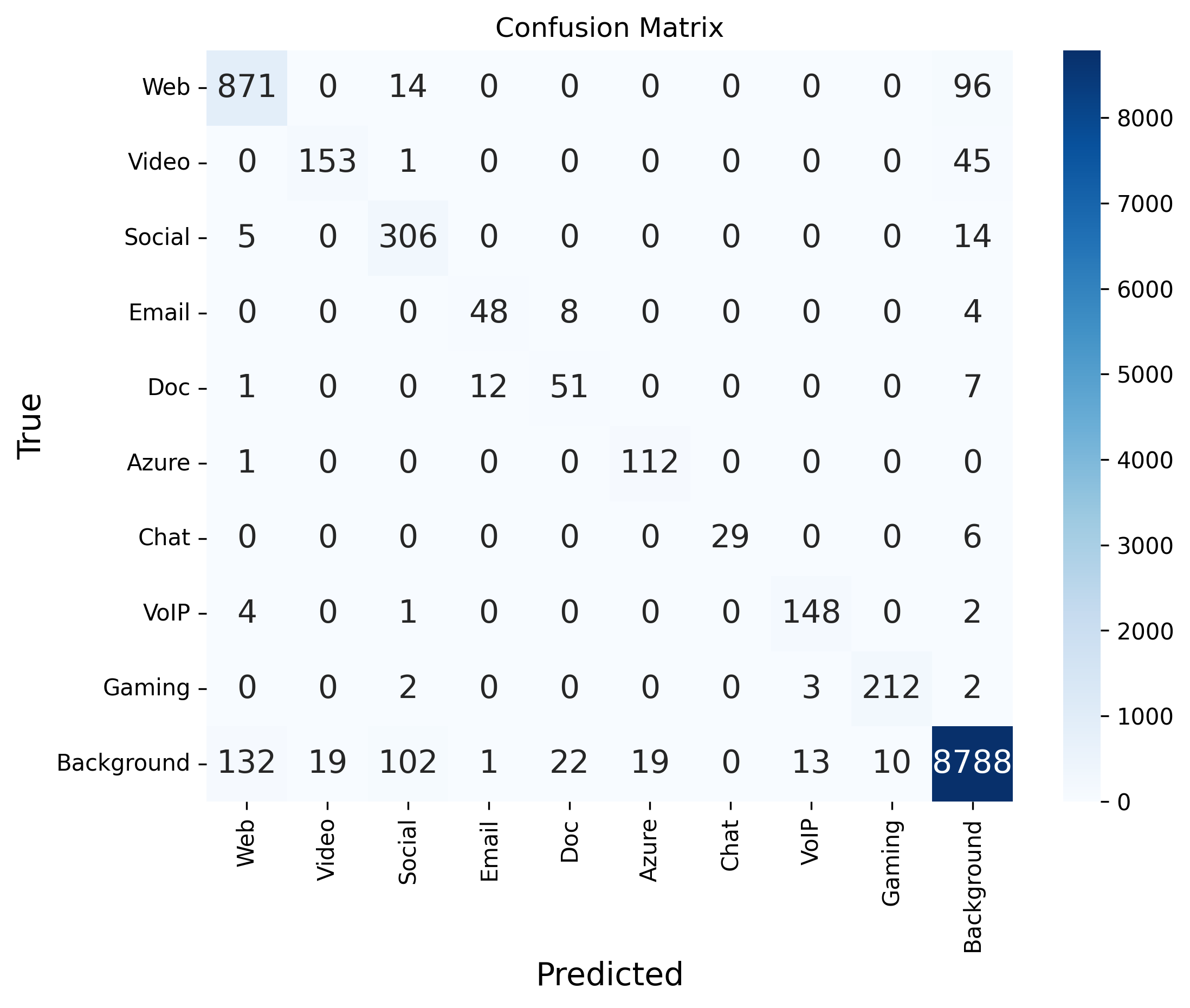}
  \caption{Comprehensively Labeled Experiment (FS-Net)}
  \label{fig:10class}
\end{figure}

We find that when we include background traffic for classification, average accuracy and weighted F1 score still remain high, as it is 0.94 in BiLSTM and 0.95 in FS-Net. However, there is notable confusion between relevant and background traffic (Fig. \ref{fig:10class}). This introduces relatively high number of false positives compared to true positives in the relevant application types, which causes the macro average F1 score to drop to 0.84 in BiLSTM and 0.87 in FS-Net. Since the number of false positive samples (in the bottom row of Fig. \ref{fig:10class}) is notably large, analyzing the performance of different applications based on such classification result would not be reliable because background traffic would be included in the analysis. Therefore, we aim to minimize the number of false positives among all samples predicted in that class.

\section{Measuring the Uncertainty}
The first question we ask is: Why does introduction of the background traffic reduce the classification performance? In this regard, the primary difference between background traffic and other relevant application traffic is that the former differs widely across application sessions and comes from a myriad of content delivery networks. As a result, there are two factors that may contribute to confusion between the relevant application and background traffic, as seen in the last row and column of Fig. \ref{fig:10class}. First, there is larger volume and greater heterogeneity of background traffic, which increases the likelihood of possible feature-space overlap with the relevant application traffic, causing background traffic to be misclassified as relevant application and vice versa. Second, by splitting the dataset by application sessions, the test sessions contain unique background traffic that differs from the background traffic seen in train sessions, causing such traffic to be misclassified as relevant application traffic. Since the deep learning model is forced to classify such uncertain traffic, this drops the performance of the model. 

One possible way to address this is to allow the classification framework to refrain from classifying traffic for which it is not certain of what the actual class is. And, one simple approach to do so is to use the softmax probability that the models outputs as a proxy for uncertainty, as is done in \cite{softmax}. That is, if the highest softmax probability across different classes is lower than some threshold, then the sample is seen as uncertain. 
\begin{figure}[h]
  \centering
  \centering
  \includegraphics[width=0.8\linewidth]{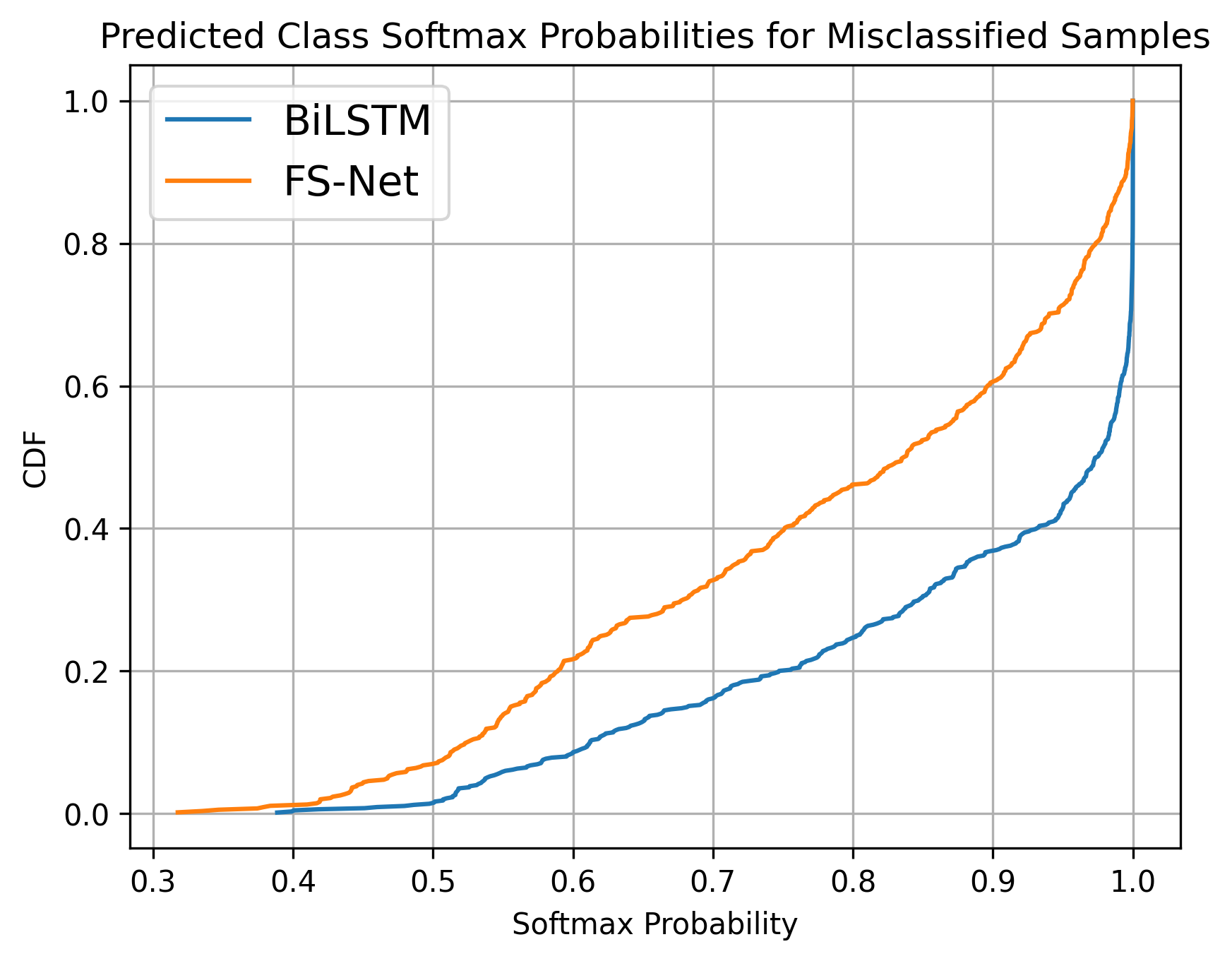}
  \caption{CDF of Predicted Class Softmax Probabilities in Misclassified Samples (BiLSTM and FS-Net)}
  \label{fig:softmax_cdf}
\end{figure}

Fig. \ref{fig:softmax_cdf} plots, for misclassified data samples, the distribution of softmax probability from the predicted class (the class with the highest probability is what the classifier returns as the label). It is natural to expect that a good classifier should have a small softmax probability for the misclassified samples--that is, it should not be very certain about the incorrect label it returns. However, Fig. \ref{fig:softmax_cdf} contradicts this. We also observe that FS-Net produces relatively lower softmax probabilities for the predicted class than our BiLSTM model.\footnote{This is likely caused by the L2 regularization on FS‑Net’s output layer, which reduces the magnitude of the logits and subsequently produces lower softmax probabilities for the predicted class.} Nevertheless, in both models, a large number of misclassified samples still have high softmax probabilities. For example, setting a threshold of 0.9 would still misclassify around 60\% and 40\% of the misclassified samples in BiLSTM and FS-Net, respectively.   

\begin{figure}[h]
  \centering
  \centering
  \includegraphics[width=0.8\linewidth]{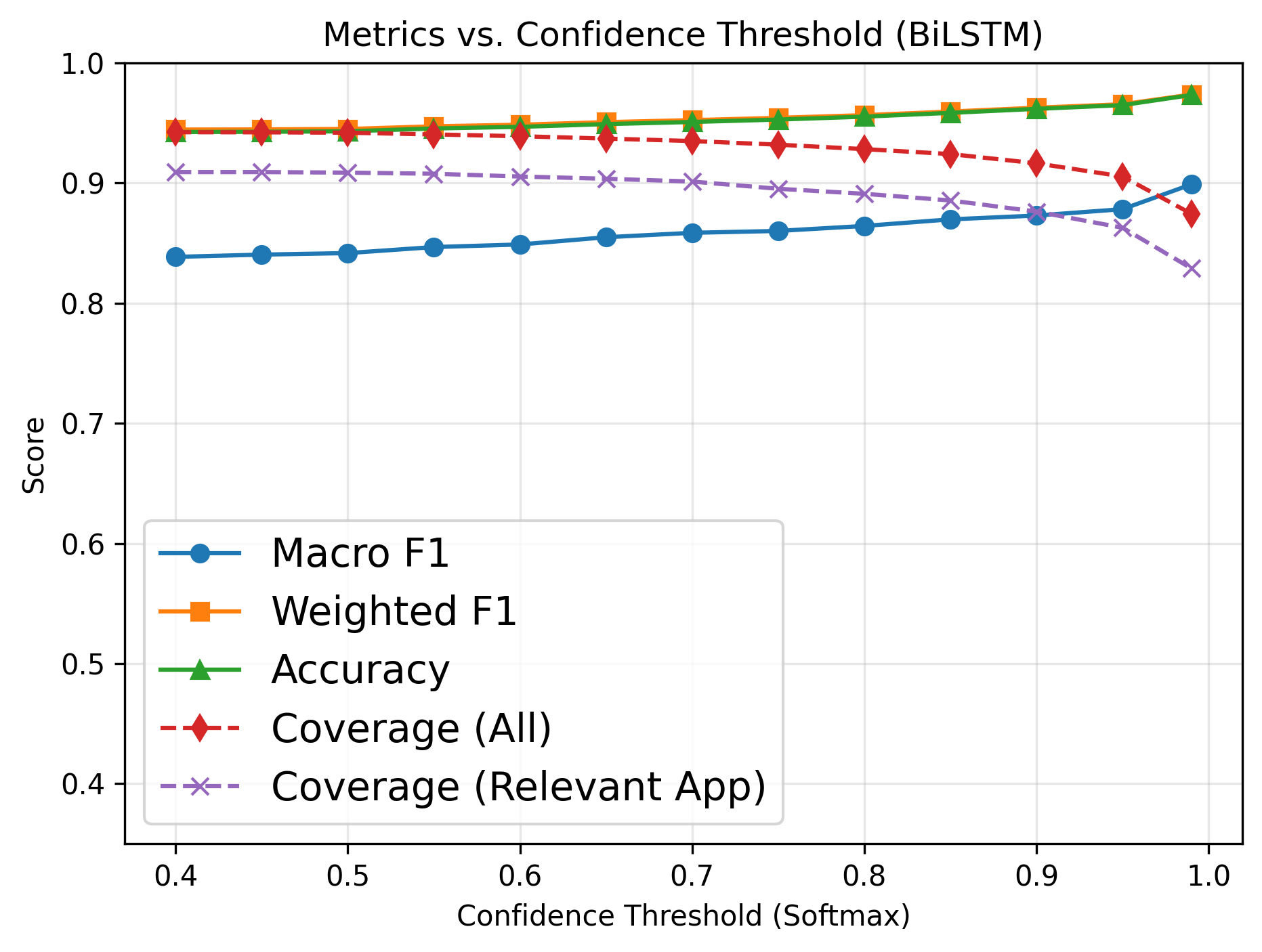}
  \caption{BiLSTM: Impact of Different Softmax Probability Thresholds}
  \label{fig:bilstm_softmax}
\end{figure}

\begin{figure}[h]
  \centering
  \centering
  \includegraphics[width=0.8\linewidth]{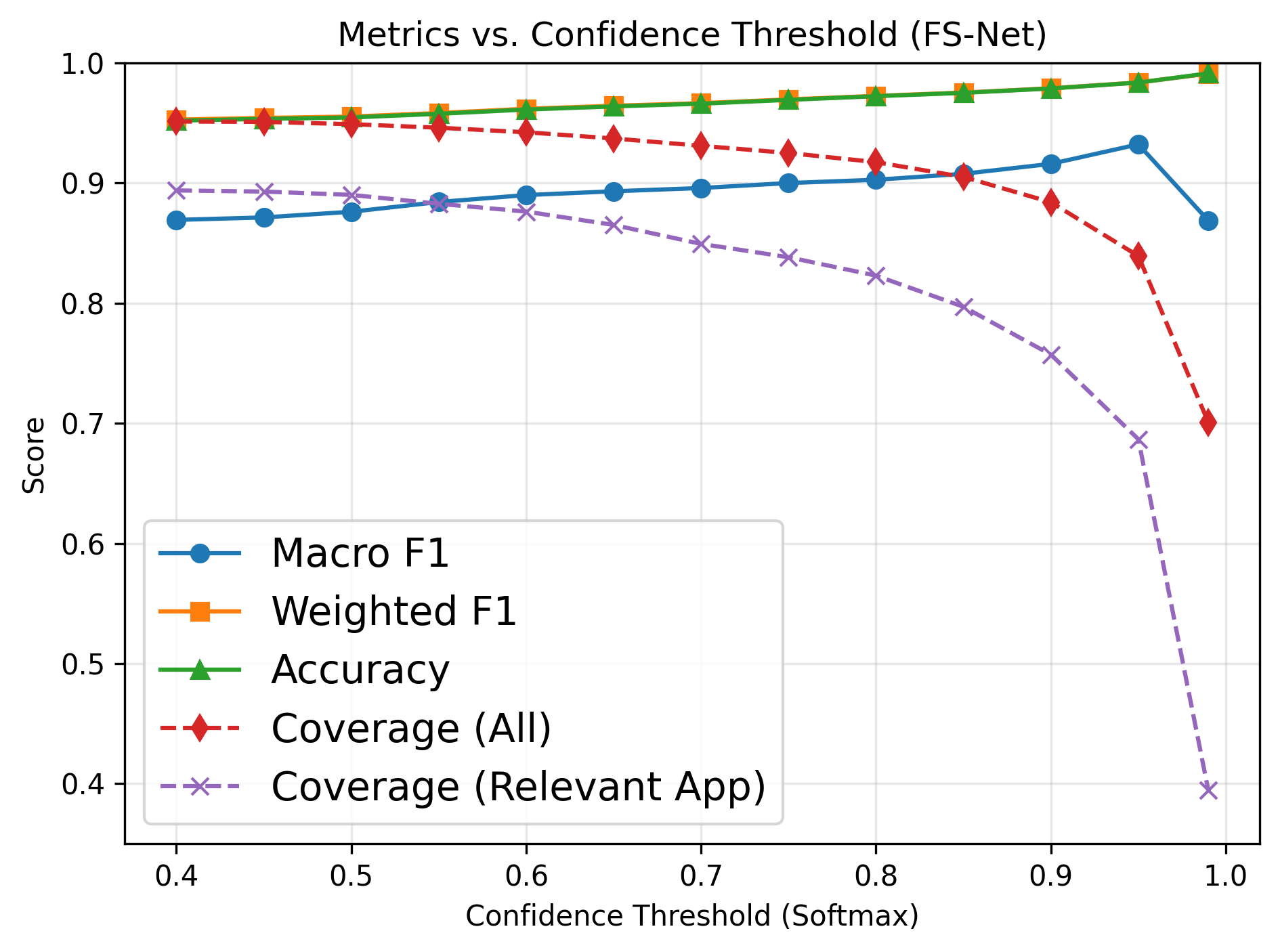}
  \caption{FS-Net: Impact of Different Softmax Probability Thresholds}
  \label{fig:fsnet_softmax}
\end{figure}

Note that using the softmax probability to filter out traffic based on uncertainty may also filter out traffic that would have been otherwise correctly labeled by the classifier. To characterize the implied trade-off between misclassification rate and coverage of correctly classified samples, we define two additional metrics: overall coverage and relevant coverage. Overall coverage is the fraction of correctly classified samples out of the total initial samples across all classes, whereas relevant coverage is the fraction of correctly classified relevant application samples out of the total initial relevant application samples (excluding background samples). Filtering samples based on the confidence threshold may cause both overall and relevant coverage to drop. In Figs. \ref{fig:bilstm_softmax} and \ref{fig:fsnet_softmax}, we plot the performance metrics when traffic is not assigned a label for softmax probabilities less than a threshold, for different softmax probability thresholds ranging from 0.4 to 0.95 at an interval of 0.05 and also include 0.99 to observe the model performance at a stringent threshold.

From Figs. \ref{fig:bilstm_softmax} and \ref{fig:fsnet_softmax}, we see improvement in the macro average F1 score for both models, with FS-Net achieving 0.93 at threshold $\geq$0.95 and BiLSTM achieving 0.90 at the threshold $\geq$0.99. Nevertheless, both BiLSTM and FS-Net lose more coverage in relevant application traffic compared to improvement in macro average F1 score, with FS-Net in particular suffering from significantly low relevant application coverage of 0.69 at $\geq$0.95. This indicates that softmax probabilities of the predicted class should be lower for uncertain samples. 

In principle, there are two forms of uncertainty that can be encountered \cite{pearce}:
\begin{enumerate}
  \item \textbf{Aleatoric uncertainty}: the uncertainty arising from different classes exhibiting overlapping feature distributions.
  \item \textbf{Epistemic uncertainty}: the uncertainty originating from the model’s limited knowledge of the input.
\end{enumerate}
However, deep learning models trained with cross-entropy loss often yield overconfident softmax probabilities, especially for samples exhibiting epistemic uncertainty \cite{pearce}. Since our dataset contains diverse traffic (especially for background), we aim to develop a classification framework that can accurately quantify its overall uncertainty.

\section{Gaussian Mixture Model Classification Framework}
Using Gaussian Mixture Model (GMM) clustering to determine how far a sample lies from a known cluster is more reliable mean to assess the epistemic uncertainty \cite{pearce}. However, in order to effectively cluster the samples, the raw time series feature vectors need to be first encoded into learned representation vector using deep learning. Prior work shows that supervised contrastive loss produces representation vectors that are better for classification than the output logits learned through cross‑entropy loss \cite{supervisedcontrastive}. Therefore, we use the supervised contrastive loss \cite{supervisedcontrastive} to train the deep learning encoder. To make the BiLSTM model compatible with the supervised contrastive loss, we remove the softmax output layer from the original architecture such that it outputs a 64-dimensional embedding vector. For FS-Net, the original loss is the sum of classification cross-entropy loss and reconstruction cross-entropy loss. We modify the original loss by replacing the classification loss with supervised contrastive loss. Also, we change the final layer of FS-Net to output a 64-dimensional embedding vector.
\subsection{Feature Vector Construction}
After training the encoder, the embedding generated by the encoder needs to be used for classification. Instead of directly using the high dimensional (64-dim) embedding for clustering, we adopt an approach from an anomaly detection framework (AOC-IDS) that operates on cosine similarity score \cite{aoc-ids}. In AOC-IDS, the average embedding of the normal traffic is computed, and the cosine similarity between the average and all of the embedding samples is fit to two Gaussian distributions (normal and abnormal traffic) using maximum likelihood estimation. During the binary classification, the test sample's cosine similarity score is compared against the two distributions and labeled according to the distribution with a higher probability. 

In our case, the encoder outputs similar embeddings for the same class, so we can characterize each sample based on how similar it is to the average embedding (centroid) of each class. Since our problem involves multiclass classification, we compute an average embedding for each of the ten classes in the training dataset. Then for a given sample, we compute its cosine similarity score to the 10 different average embeddings, which yields a compact 10-dimensional cosine similarity score vector. 
\subsection{GMM Clustering}
We fit the GMM to our training dataset by using the Expectation-Maximization algorithm \cite{EM}. After 10 separate clusters are formed, we label each cluster based on the majority of the samples that end up in each cluster.

\begin{table}[h]
\centering
\begin{tabular}{ll}
\toprule
\textbf{FS‑Net} & \textbf{BiLSTM} \\
\midrule

{\bf Web Browsing}: 24056 (98.81\%)  
  & {\bf Web Browsing}: 23,248 (99.60\%) \\
Background:   289 (1.19\%)  
  & Background:    85 (0.36\%) \\
                                    & Social Media:   6  (0.03\%) \\
                                    & Video:          2  (0.01\%) \\
\addlinespace[0.5em]
\midrule

{\bf Video}:      24535 (99.93\%)  
  & {\bf Video}:      23,484 (99.92\%) \\
Background:     17 (0.07\%) 
  & Background:     19 (0.08\%) \\
\addlinespace[0.5em]
\midrule

{\bf Social Media}:   24535 (99.79\%)  
  & {\bf Social Media}: 23,917 (99.84\%) \\
Background:        52 (0.21\%)
  & Background:      25 (0.10\%) \\
  & Web Browsing:    12 (0.05\%) \\
                                    & Video:            1  (0.00\%) \\
\addlinespace[0.5em]
\midrule

{\bf Email}:      24535 (99.75\%)  
  & {\bf Email}:      23,729 (99.92\%) \\
Background:    61 (0.25\%)
  & Background:    12 (0.05\%) \\
                                    & Online Docs:     6  (0.03\%) \\
                                    & Web Browsing:    1  (0.00\%) \\
\addlinespace[0.5em]
\midrule

{\bf Online Docs}: 24535 (99.96\%)  
  & {\bf Online Docs}: 23,835 (99.95\%) \\
Background:    10 (0.04\%)
  & Email:          7  (0.03\%) \\
                                    & Background:     4  (0.02\%) \\
\addlinespace[0.5em]
\midrule

{\bf Microsoft Azure}: 24535 (99.91\%)  
  & {\bf Microsoft Azure}: 23,616 (99.99\%) \\
Background:       21 (0.09\%) 
  & Background:       3  (0.01\%) \\
\addlinespace[0.5em]
\midrule

{\bf Chat}:         24535 (99.98\%) 
  & {\bf Chat}:         24,025 (99.99\%) \\
Background:      4 (0.02\%) 
  & Background:      3  (0.01\%) \\
\addlinespace[0.5em]
\midrule

{\bf VoIP}:         24535 (99.93\%) 
  & {\bf VoIP}:         24,441 (99.96\%) \\
Background:     18 (0.07\%) 
  & Background:     9  (0.04\%) \\
\addlinespace[0.5em]
\midrule

{\bf Gaming}:       24535 (99.95\%)  
  & {\bf Gaming}:       23,691 (99.99\%) \\
Background:     13 (0.05\%) 
  & Background:     2  (0.01\%) \\
\addlinespace[0.5em]
\midrule

{\bf Background}: 24050 (98.05\%)  
  & {\bf Background}: 24,373 (78.20\%) \\
Web Browsing: 479 (1.95\%)  
  & Web Browsing: 1,274  (4.09\%) \\
  & Video:        1,048  (3.36\%) \\
                                    & Microsoft Azure: 919  (2.95\%) \\
                                    & Gaming:      844  (2.71\%) \\
                                    & Email:       799  (2.56\%) \\
                                    & Online Docs: 694  (2.23\%) \\
                                    & Social Media:612  (1.96\%) \\
                                    & Chat:        510  (1.64\%) \\
                                    & VoIP:         94  (0.30\%) \\
\bottomrule
\end{tabular}
\caption{Cluster Analysis: FS‑Net vs. BiLSTM True‑Label Breakdown}
\label{tab:clusters}
\end{table}

Table \ref{tab:clusters} compares the 10 different clusters formed by FS-Net and BiLSTM encoder from the training dataset. Cosine similarity score vectors derived from both encoders cluster relevant application traffic quite effectively (with around 99\% of each cluster composed of a unique application type), with a better background cluster as well when using the FS-Net encoder. This suggests that the combination of our supervised contrastive loss encoders, and GMM clustering based on cosine similarities of the embeddings, does an excellent job of identifying the similarities and differences between different classes in the training dataset. For classification of test samples, we compute the cosine similarity between each sample's embedding and the ten average embeddings computed from the training dataset. Afterward, we label it based on the GMM cluster to which it belongs. 

To evaluate the impact of the 10-dimensional cosine similarity score vectors, we also run experiments where we cluster the 64-dimensional embedding vectors directly from the same BiLSTM and FS-Net encoders. Table \ref{tab:gmmresult} shows the result for the four different classification evaluations.
\begin{table}[H]
  \centering
  \scalebox{1}{%
  \begin{tabular}{lccc}
    \hline
     \textbf{Encoder} & \textbf{Macro F1} & \textbf{Accuracy} & \textbf{Weighted F1} \\
    \hline
     BiLSTM (Cosine Sim) &0.85     & 0.95     & 0.95         \\
     \hline
     BiLSTM (Embedding) &0.81     & 0.87     & 0.88         \\
     \hline
    FS-Net (Cosine Sim) &0.86 & 0.95 & 0.96\\
    \hline
    FS-Net (Embedding) &0.85 & 0.95 & 0.95\\
    \hline
  \end{tabular}
  }
    \caption{GMM Classification Performance}
  \label{tab:gmmresult}
\end{table}
We find that using cosine similarity scores instead of the 64-dimensional embedding vectors for the GMM clustering improves the results for both encoders, with significant improvement for BiLSTM encoder. Furthermore, this classification framework achieves comparable classification performance to the same model that was trained with cross-entropy loss for classification (Table \ref{tab:comprehensive}), with macro average F1 score improving from 0.84 to 0.85 for BiLSTM and decreasing from 0.87 to 0.86 for FS-Net.
\subsection{Characterizing Uncertainty}
\begin{figure}[h]
    \centering
    \includegraphics[width=0.9\linewidth]{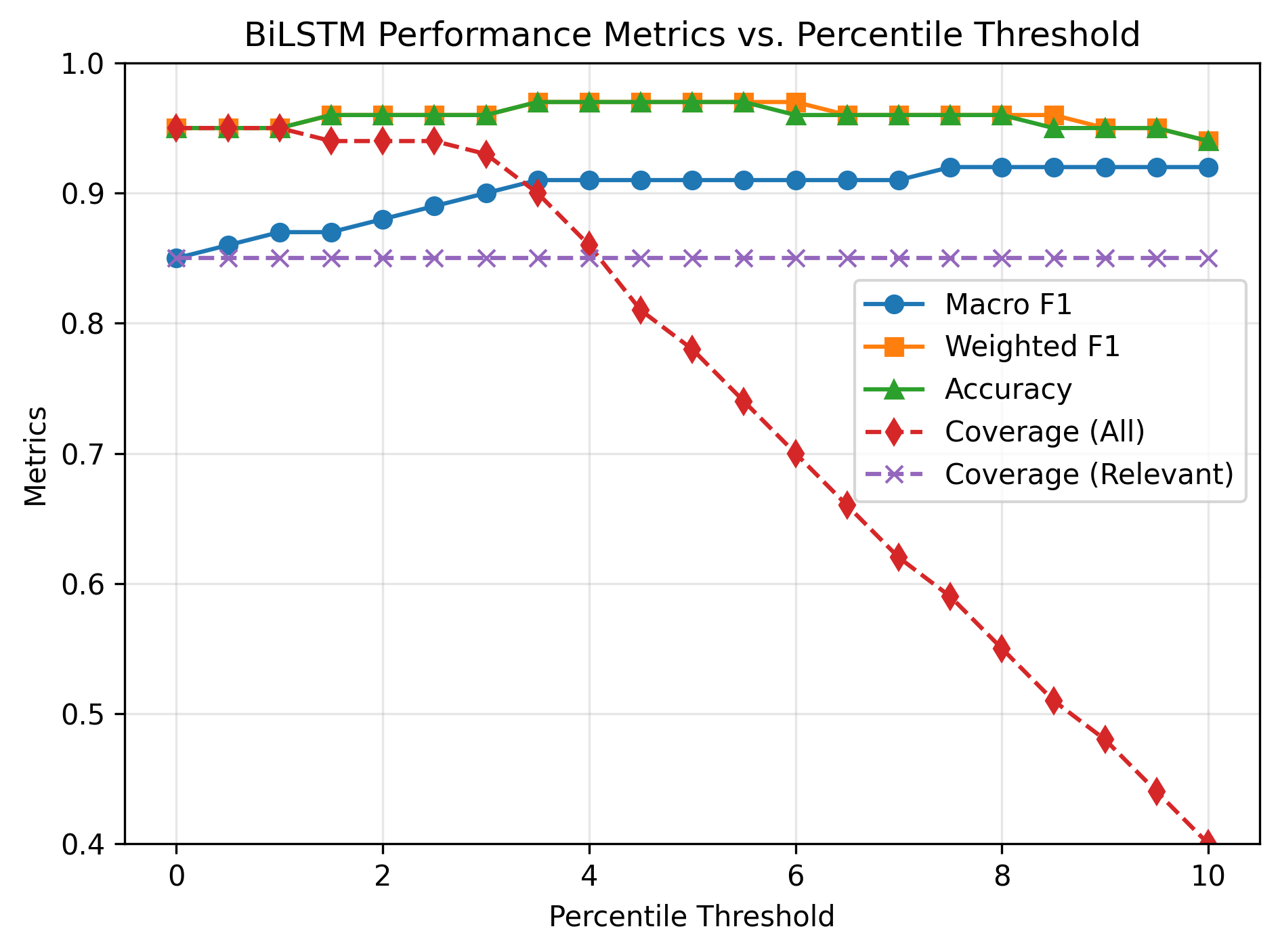}
    \caption{Different Percentile Thresholds (BiLSTM-GMM)}
  \label{fig:bilstmgmm}
\end{figure}

\begin{figure}[h]
  \centering
    \includegraphics[width=0.9\linewidth]{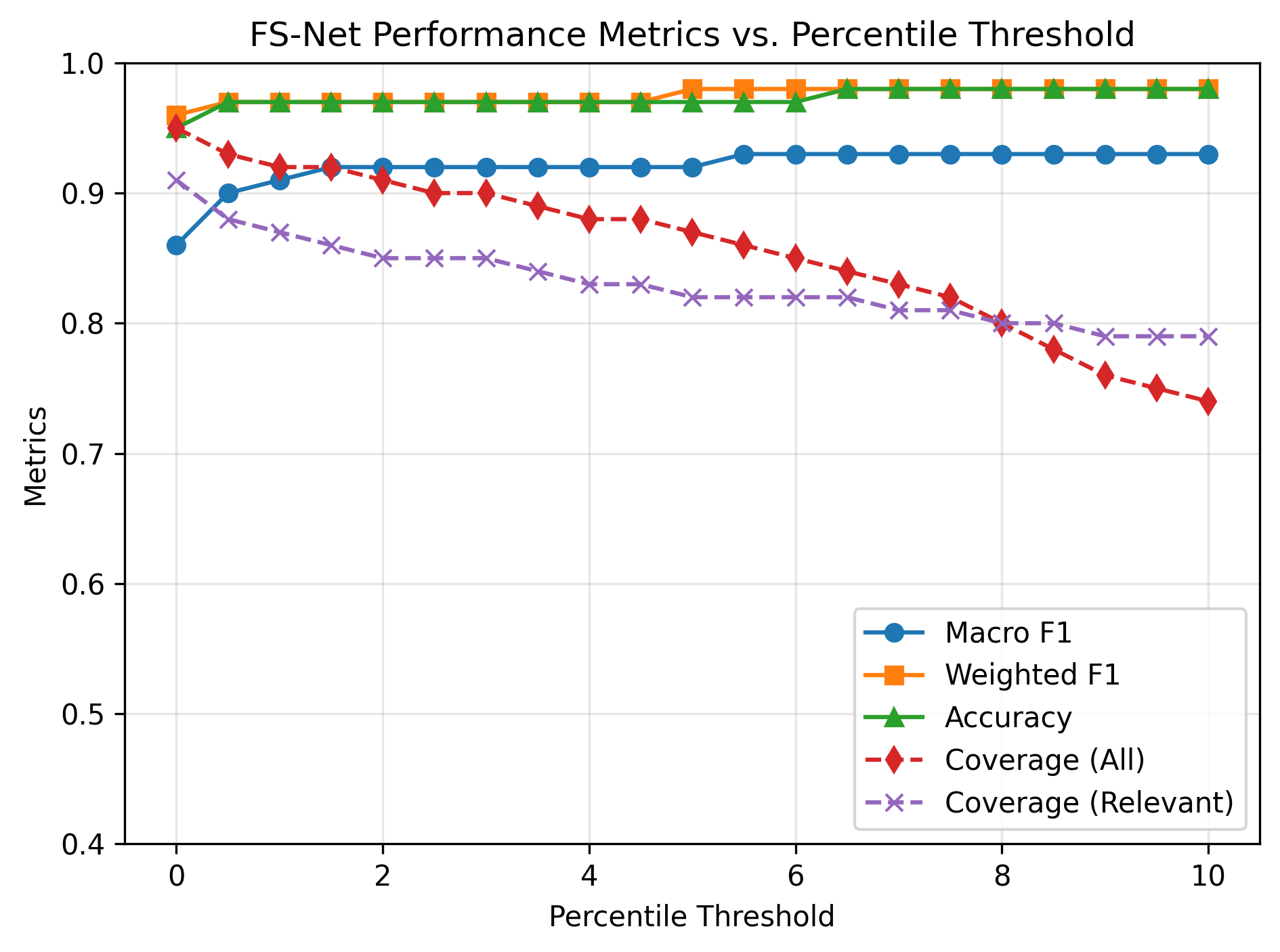}
    \caption{Different Percentile Thresholds (FS‑Net-GMM)}
  \label{fig:fsnetgmm}
\end{figure}

Still, the main goal of this method is to improve the assessment of the confidence to eliminate samples with lower certainty. In GMM, outliers are typically identified by computing each training sample’s log‑likelihood, choosing a threshold at a specified percentile of those values, and then labeling any sample with a log‑likelihood below that threshold as an outlier \cite{gmmfilter}. We adapt this approach to filter uncertain samples from classification. To examine the trade-off between reliable classification and coverage, we vary the percentile threshold from a reasonable range of 0 to 10 at an interval of 0.5. Figs. \ref{fig:bilstmgmm} and \ref{fig:fsnetgmm} show the performance across different metrics for GMM using FS-Net encoder (FS-Net-GMM) and BiLSTM encoder (BiLSTM-GMM). 

As seen in Fig. \ref{fig:bilstmgmm}, for BiLSTM-GMM, we see a steady increase in the macro average F1 score with only the overall coverage impacted by the increase in the percentile threshold. The relevant application coverage remains the same because in this GMM, many background traffic samples are assigned lower log-likelihood values than correctly identified relevant application traffic samples, causing mostly background traffic samples to be filtered based on the threshold. Based on Fig. \ref{fig:fsnetgmm}, for FS-Net-GMM, we observe that the macro average F1 score improves notably at 0.5th percentile, and as the percentile threshold increases, there is a steady increase in the macro average F1 score at the expense of a steady decrease in both coverages with the overall coverage decreasing at a faster rate than the relevant application coverage. 




It is important to highlight that using log-likelihood percentile threshold (within our GMM-based classification framework) aligns more with the intuitive expectation for confidence threshold, since increasing the percentile threshold improves the macro average F1 score at the expense of steady coverage loss. In contrast, we observed different trends when using softmax probabilities as the basis for uncertainty. For example, for FS-Net, a lot of correctly classified samples also had lower softmax probability than the specified threshold, causing substantial coverage loss without much improvement in performance at higher end of the threshold (Fig. \ref{fig:fsnet_softmax}). In addition, for BiLSTM, the maximum macro average F1 score achieved at the stringent softmax threshold of $\geq$ 0.99 was 0.9 (Fig. \ref{fig:bilstm_softmax}), indicating that many misclassified samples also had excessively high softmax probabilities. 

We report below the notable thresholds where BiLSTM-GMM and FS-Net-GMM achieved the best coverage for macro average F1 score $\geq$ 0.90 in Table \ref{tab:gmmperf}. As seen in Table \ref{tab:gmmperf}, when considering the overall and relevant application coverage, 0.5th percentile threshold for FS-Net-GMM and 3rd percentile for BiLSTM-GMM are reasonable. However, if the focus is more on reliably classifying the relevant traffic, it may be reasonable to have a higher percentile threshold such as 5.5th percentile for FS-Net-GMM and 7.5th percentile for BiLSTM-GMM. 
\begin{table}[H]
\centering
\scriptsize                   
\begin{tabular}{lcccc}
\hline
\textbf{Model}   & \textbf{Threshold} & \textbf{Macro F1} & \textbf{Coverage (All)} & \textbf{Coverage (Relevant)} \\
\hline
BiLSTM  & 3       & 0.90    & 0.93           & 0.85                 \\
\hline
BiLSTM  & 3.5       & 0.91    & 0.86           & 0.85                 \\
\hline
BiLSTM  & 7.5       & 0.92    & 0.59           & 0.85                 \\
\hline
FS-Net  & 0.5       & 0.90    & 0.93           & 0.88                 \\
\hline
FS-Net  & 1       & 0.91    & 0.92           & 0.87                 \\
\hline
FS-Net  & 1.5       & 0.92    & 0.92           & 0.86                 \\
\hline
FS-Net  & 5.5       & 0.93    & 0.86           & 0.82                 \\
\hline
\end{tabular}
\caption{Performance at Specific Percentile Threshold}
\label{tab:gmmperf}
\end{table}

\begin{figure}[h]
  \centering
  \includegraphics[width=0.8\linewidth]{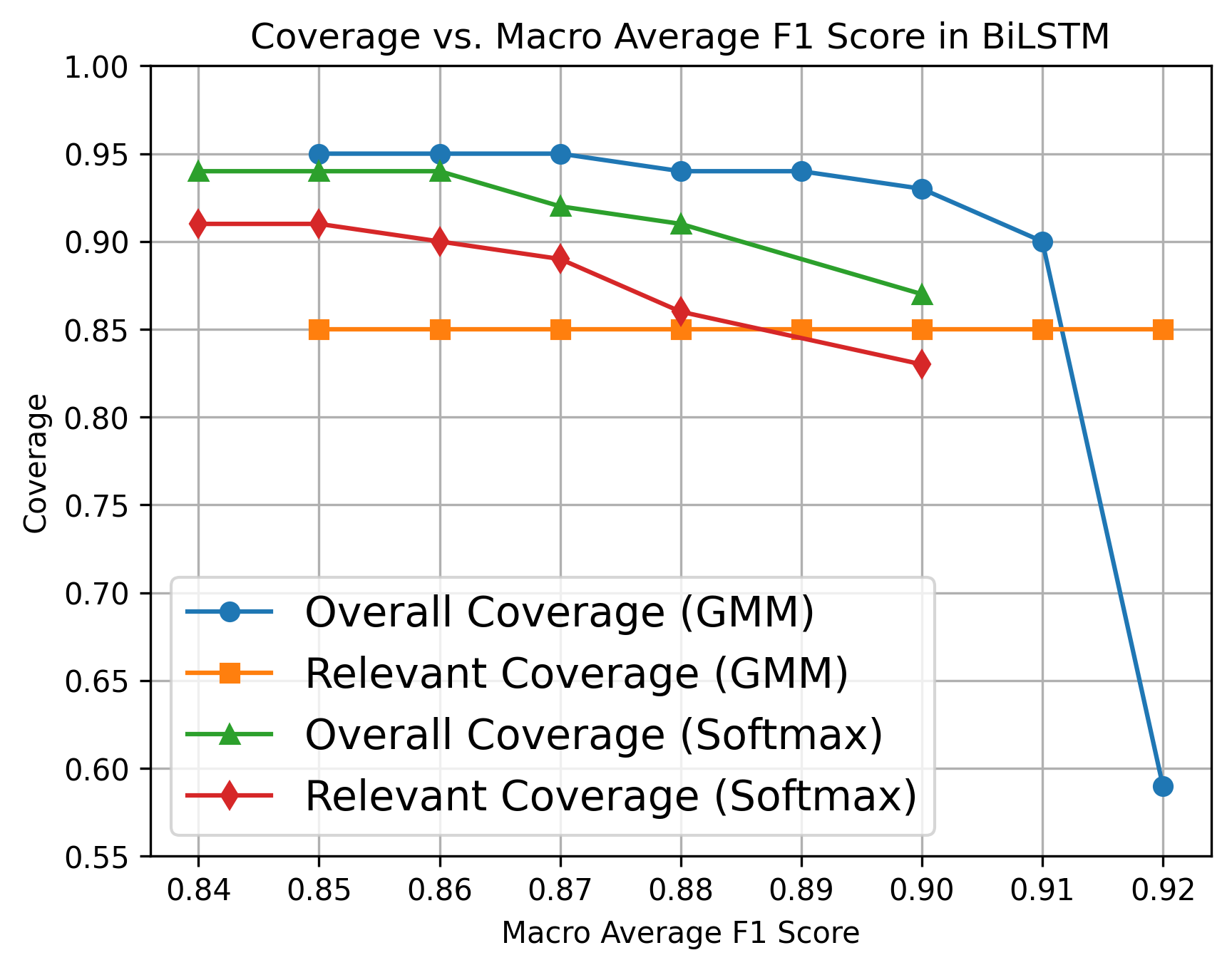}
  \caption{Macro Average F1 Score vs. Coverage (BiLSTM)}
  \label{fig:bilstmcoverage}
\end{figure}

\begin{figure}[h]
  \centering
    \includegraphics[width=0.8\linewidth]{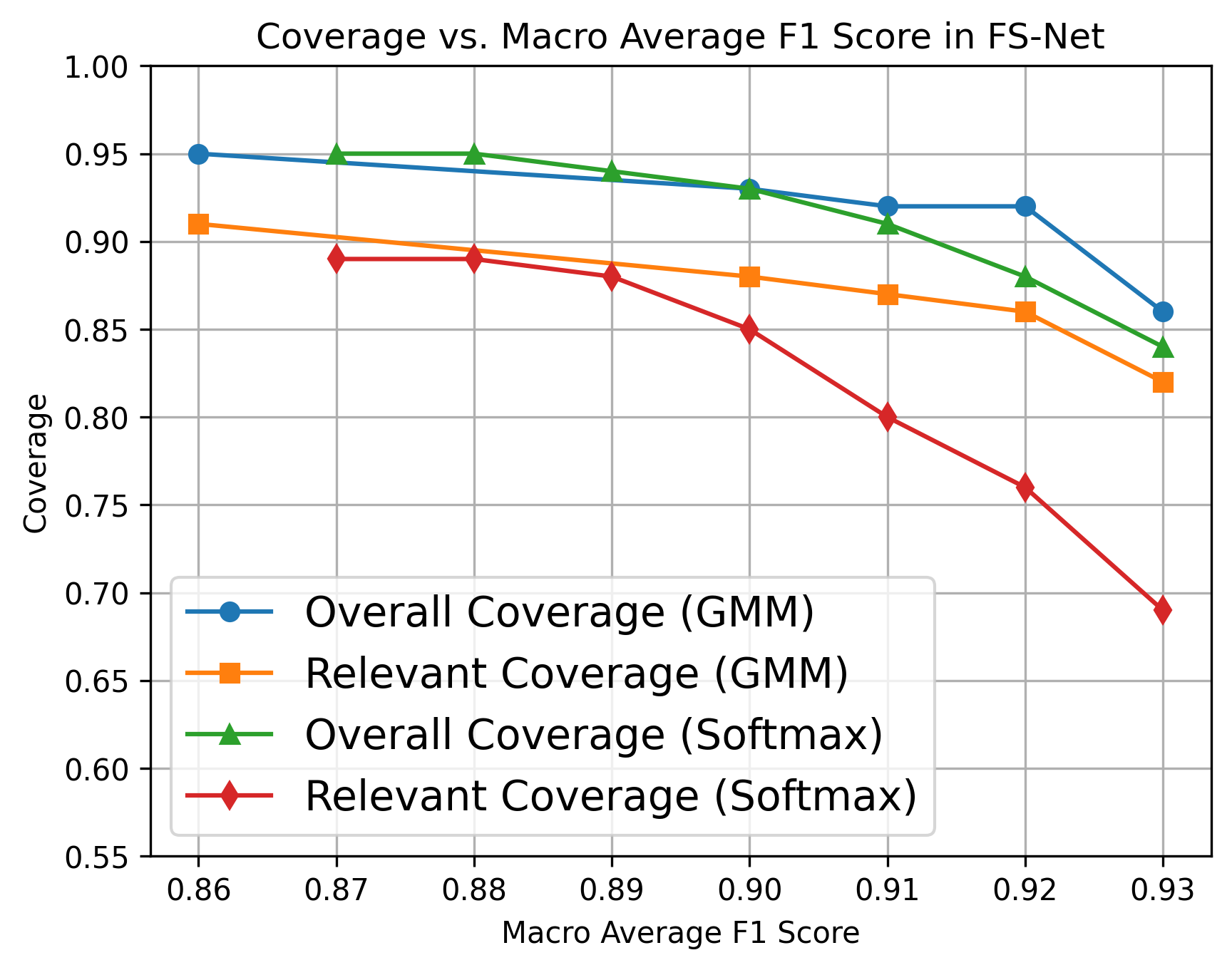}
    \caption{Macro Average F1 Score vs. Coverage (FS‑Net)}
  \label{fig:fsnetcoverage}
\end{figure}

More fundamentally, our classification framework gives better coverage than softmax probability threshold (0.4-0.99) for different macro average F1 scores. As seen from Figs. \ref{fig:bilstmcoverage} and \ref{fig:fsnetcoverage}, for macro average F1 score $\geq$0.9, both overall and relevant coverage are mostly higher when using GMM log-likelihood percentile threshold as basis for uncertainty in comparison to softmax probability threshold. For BiLSTM-GMM, this percentile threshold allows macro average F1 score to reach 0.92 without substantial loss in relevant application coverage. For FS-Net-GMM, this percentile threshold allows significantly higher relevant application coverage than the softmax threshold, demonstrating its effectiveness to filter uncertain samples without incurring too much expense in ignoring correctly classified relevant application samples.

\begin{figure}[H]
  \centering
  \centering
  \includegraphics[width=0.9\linewidth]{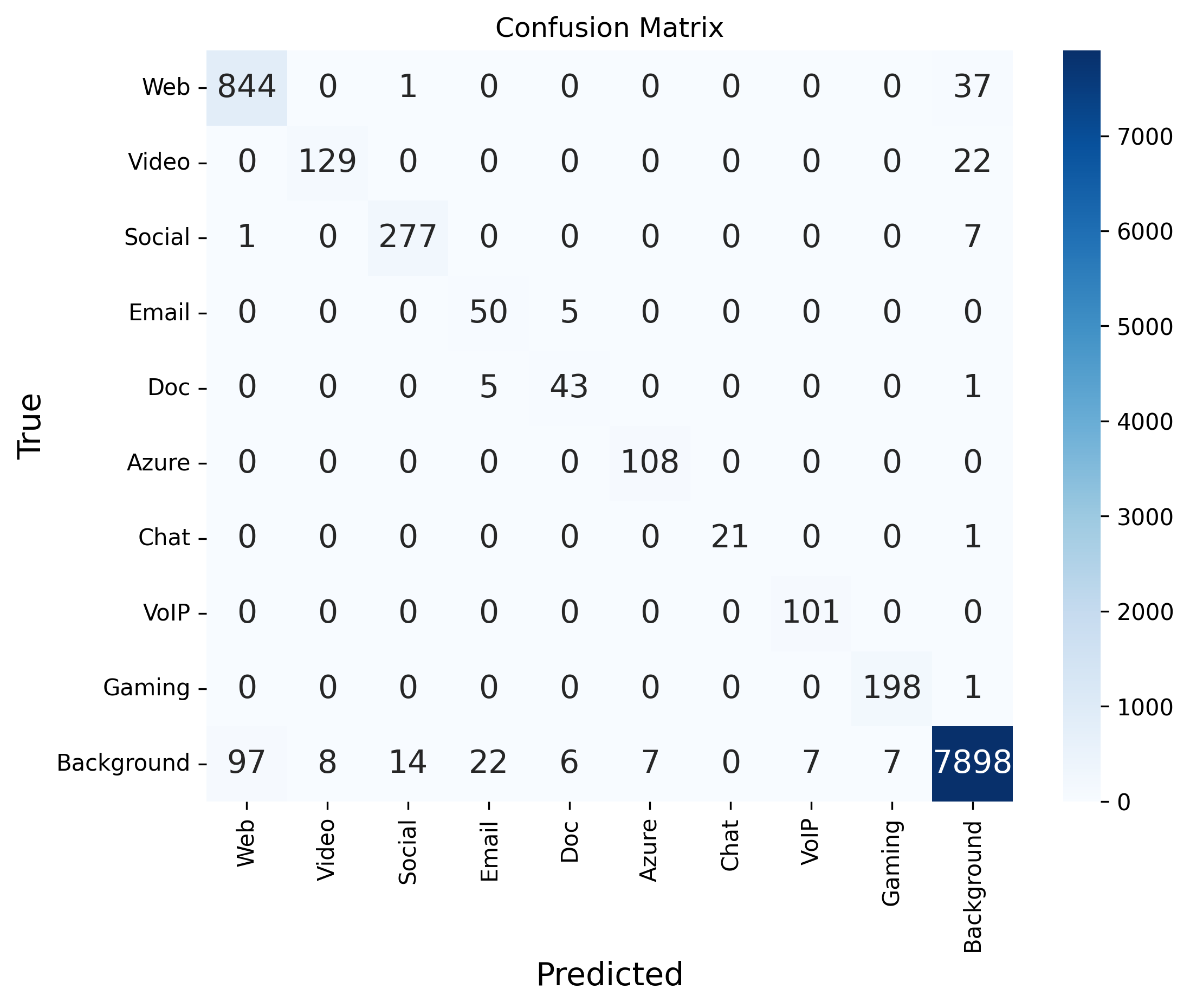}
  \caption{FS-Net-GMM Confusion Matrix (5.5th Percentile Threshold)}
  \label{fig:gmmfsnet}
\end{figure}

We show FS-NET-GMM’s classification result at one of the notable threshold (5.5th) in Fig. \ref{fig:gmmfsnet}. The bottom row of the confusion matrix reveals a notable reduction in background flows that are misclassified as relevant application traffic compared to Fig. \ref{fig:10class}. In particular, background to social media errors noticeably decreased from 102 to just 14 samples. Despite the overall improvements, many background traffic remained misclassified as web browsing likely due to inherent overlap of their feature-space distributions, which FS-NET-GMM cannot reliably distinguish without incurring too much coverage loss. Still, compared to the softmax threshold baseline, the proposed framework detects uncertain samples better without substantial coverage loss, making it more applicable for reliably classifying application traffic in the presence of generic background traffic.

\section{Related Work}
One of the reference model used in this paper is Flow Sequence Network (FS-Net) \cite{fs}. In FS-Net, packet size sequences are first passed through a small embedding layer, producing a learnable vector at each time
step. The embeddings are then accepted by the GRU-based encoder and then subsequently fed into a GRU-based decoder. For classification, FS-Net combines the output of encoder and decoder and passes it through a dense layer to produce logits, which are used in the cross-entropy loss. The decoder’s output sequence is also compared against the original inputs to account for the reconstruction loss. The model is trained by minimizing the sum of the classification cross-entropy loss and the reconstruction cross-entropy loss. 

The time series features we used for our BiLSTM model is derived from a recent study \cite{akbari}. In their study, they use three broad categories of features: unencrypted TLS handshake header bytes (with SNI and cipher info removed), flow time series (packet sizes, direction, and inter-arrival times), and flow statistics (such as mean/median/standard deviations of  packet sizes, number of distinct TCP flags, and flow duration). The three inputs are fed in three separate neural network layers and subsequently fed into the same fully connected layers to produce the final outcome. The TLS handshake header is fed into 1D CNN with max pooling layers, flow time series are fed into dense layer followed by three LSTM layers, and flow statistics are fed into fully connected layers. 

From another study AOC-IDS \cite{aoc-ids}, we adopt the conversion of embedding into cosine similarity score to use for classification. They compute the cosine similarity between each sample and the average representation vector of the normal traffic, producing a similarity score per sample. Then, they fit two Gaussian distributions to these scores: one over the similarity scores of normal samples and the other over that of abnormal samples. For classification, the similarity score of the sample is compared against these two Gaussian distributions to be labeled according to the distribution with higher probability.
\section{Conclusion}
Accurately classifying application traffic is essential to monitor the usage, performance, and resource needs of different applications. State-of-the-art application classifiers, however, are evaluated on curated datasets that disregard background traffic, such as advertisements, analytics, shared APIs, and trackers that are associated with the applications. Including such background traffic for classification causes overall performance loss in the model due to notable confusion between relevant application and background traffic. Therefore, there needs to be a reliable confidence measure that limits the predictions to only the most certain samples. However, in line with previous research, we find that the deep learning model’s softmax outputs are unreliable, as they tend to overestimate the confidence of the uncertain samples. To address this, we propose a GMM-based classification framework that more accurately filters the uncertain samples, preserving greater traffic coverage compared to the softmax threshold baseline. A promising direction for future work is to extend this application classification framework to an online setting, where the proposed confidence measure can be used to detect concept drift (based on the proportion of uncertain samples) and trigger model retraining when necessary.

\printbibliography

@article{akbari,
author = {Akbari, Iman and Salahuddin, Mohammad A. and Aniva, Leni and Limam, Noura and Boutaba, Raouf and Mathieu, Bertrand and Moteau, Stephanie and Tuffin, Stephane},
title = {Traffic classification in an increasingly encrypted web},
year = {2022},
issue_date = {October 2022},
publisher = {Association for Computing Machinery},
address = {New York, NY, USA},
volume = {65},
number = {10},
issn = {0001-0782},
doi = {10.1145/3559439},
journal = {Commun. ACM},
month = sep,
pages = {75–83},
numpages = {9}
}

@inproceedings{etbert,
  author    = {Xinjie Lin and
               Gang Xiong and
               Gaopeng Gou and
               Zhen Li and
               Junzheng Shi and
               Jing Yu},
  title     = {{ET-BERT:} {A} Contextualized Datagram Representation with Pre-training
               Transformers for Encrypted Traffic Classification},
  booktitle = {{WWW} '22: The {ACM} Web Conference 2022, Virtual Event, Lyon, France,
               April 25 - 29, 2022},
  pages     = {633--642},
  publisher = {{ACM}},
  year      = {2022}
}

@misc{iscx,
  author       = {University of New Brunswick, Canadian Institute for Cybersecurity},
  title        = {VPN DataSets},
  year         = {n.d.},
  url          = {https://www.unb.ca/cic/datasets/vpn.html}
}

@misc{quic,
  author       = {Guillaume Fraysse},
  title        = {{UC Davis QUIC} Dataset},
  year         = {n.d.},
  url          = {https://www.kaggle.com/datasets/guillaumefraysse/ucdavisquic}
}

@misc{sandvine2024,
  author       = {Sandvine},
  title        = {Global Internet Phenomena Report 2024},
  year         = {2024},
  howpublished = {\url{https://www.sandvine.com/global-internet-phenomena-report-2024}},
}

@misc{similarweb,
  author = {},
  title = {Top Websites Ranking},
  year = {2023},
  howpublished = {\url{https://www.similarweb.com/top-websites/}},
note = {Accessed: October 2023}
}

@misc{semrush,
  author = {{SEMrush}},
  title = {Most Searched Keywords on {Google}},
  year = {2024},
  url = {https://www.semrush.com/blog/most-searched-keywords-google/},
  note = {Accessed: October 2023}
}

@misc{popular,
  author = {Statista},
  title = {Most Viewed YouTube Videos Worldwide},
  year = {2024},
  url = {https://www.statista.com/statistics/249396/top-youtube-videos-views/},
  note = {Accessed: January 2024}
}

@article{ode,
title = {Fast and lean encrypted Internet traffic classification},
journal = {Computer Communications},
volume = {186},
pages = {166-173},
year = {2022},
issn = {0140-3664},
doi = {https://doi.org/10.1016/j.comcom.2022.02.003},
author = {Sangita Roy and Tal Shapira and Yuval Shavitt}
}

@inproceedings{augment,
  author    = {Wang, C. and Finamore, A. and Michiardi, P. and Gallo, M. and Rossi, D.},
  title     = {Data Augmentation for Traffic Classification},
  booktitle = {Passive and Active Measurement. PAM 2024},
  editor    = {Richter, P. and Bajpai, V. and Carisimo, E.},
  series    = {Lecture Notes in Computer Science},
  volume    = {14537},
  year      = {2024},
  publisher = {Springer, Cham},
  doi       = {10.1007/978-3-031-56249-5_7},
}

@misc{pearce,
      title={Understanding Softmax Confidence and Uncertainty}, 
      author={Tim Pearce and Alexandra Brintrup and Jun Zhu},
      year={2021},
      eprint={2106.04972},
      archivePrefix={arXiv},
      primaryClass={cs.LG},
}

@misc{supervisedcontrastive,
      title={Supervised Contrastive Learning}, 
      author={Prannay Khosla and Piotr Teterwak and Chen Wang and Aaron Sarna and Yonglong Tian and Phillip Isola and Aaron Maschinot and Ce Liu and Dilip Krishnan},
      year={2021},
      eprint={2004.11362},
      archivePrefix={arXiv},
      primaryClass={cs.LG},
}

@INPROCEEDINGS{aoc-ids,
  author={Zhang, Xinchen and Zhao, Running and Jiang, Zhihan and Sun, Zhicong and Ding, Yulong and Ngai, Edith C.H. and Yang, Shuang-Hua},
  booktitle={IEEE INFOCOM 2024 - IEEE Conference on Computer Communications}, 
  title={AOC-IDS: Autonomous Online Framework with Contrastive Learning for Intrusion Detection}, 
  year={2024},
  volume={},
  number={},
  pages={581-590},
  doi={10.1109/INFOCOM52122.2024.10621346}}

@ARTICLE{EM,
  author={Moon, T.K.},
  journal={IEEE Signal Processing Magazine}, 
  title={The expectation-maximization algorithm}, 
  year={1996},
  volume={13},
  number={6},
  pages={47-60},
  doi={10.1109/79.543975}}

@INPROCEEDINGS{fs,
  author={Liu, Chang and He, Longtao and Xiong, Gang and Cao, Zigang and Li, Zhen},
  booktitle={IEEE INFOCOM 2019 - IEEE Conference on Computer Communications}, 
  title={FS-Net: A Flow Sequence Network For Encrypted Traffic Classification}, 
  year={2019},
  volume={},
  number={},
  pages={1171-1179},
  doi={10.1109/INFOCOM.2019.8737507}}

@article{bilstm,
author = {Schuster, Mike and Paliwal, Kuldip},
year = {1997},
month = {12},
pages = {2673 - 2681},
title = {Bidirectional recurrent neural networks},
volume = {45},
journal = {Signal Processing, IEEE Transactions on},
doi = {10.1109/78.650093}
}

@ARTICLE{snimask,
  author={Rezaei, Shahbaz and Kroencke, Bryce and Liu, Xin},
  journal={IEEE Access}, 
  title={Large-Scale Mobile App Identification Using Deep Learning}, 
  year={2020},
  volume={8},
  number={},
  pages={348-362},
  doi={10.1109/ACCESS.2019.2962018}}

@inproceedings{rosetta,
author = {Xie, Renjie and Cao, Jiahao and Dong, Enhuan and Xu, Mingwei and Sun, Kun and Li, Qi and Shen, Licheng and Zhang, Menghao},
title = {Rosetta: enabling robust TLS encrypted traffic classification in diverse network environments with TCP-aware traffic augmentation},
year = {2023},
isbn = {978-1-939133-37-3},
publisher = {USENIX Association},
address = {USA},
booktitle = {Proceedings of the 32nd USENIX Conference on Security Symposium},
articleno = {36},
numpages = {18},
location = {Anaheim, CA, USA},
series = {SEC '23}
}

@article{gmmfilter,
  author       = {Weizun Zhao and
                  Lishuai Li and
                  Sameer Alam and
                  Yanjun Wang},
  title        = {An Incremental Clustering Method for Anomaly Detection in Flight Data},
  journal      = {CoRR},
  volume       = {abs/2005.09874},
  year         = {2020},
  eprinttype    = {arXiv},
  eprint       = {2005.09874},
  bibsource    = {dblp computer science bibliography, https://dblp.org}
}

@INPROCEEDINGS{malware,
  author={Wang, Wei and Zhu, Ming and Zeng, Xuewen and Ye, Xiaozhou and Sheng, Yiqiang},
  booktitle={2017 International Conference on Information Networking (ICOIN)}, 
  title={Malware traffic classification using convolutional neural network for representation learning}, 
  year={2017},
  volume={},
  number={},
  pages={712-717},
  doi={10.1109/ICOIN.2017.7899588}}

@INPROCEEDINGS{flowpic,
  author={Shapira, Tal and Shavitt, Yuval},
  booktitle={IEEE INFOCOM 2019 - IEEE Conference on Computer Communications Workshops (INFOCOM WKSHPS)}, 
  title={FlowPic: Encrypted Internet Traffic Classification is as Easy as Image Recognition}, 
  year={2019},
  volume={},
  number={},
  pages={680-687},
  doi={10.1109/INFCOMW.2019.8845315}}

@inproceedings{discriminative,
author = {Zhao, Lixin and Cai, Lijun and Yu, Aimin and Xu, Zhen and Meng, Dan},
title = {A novel network traffic classification approach via discriminative feature learning},
year = {2020},
isbn = {9781450368667},
publisher = {Association for Computing Machinery},
address = {New York, NY, USA},
doi = {10.1145/3341105.3373844},
booktitle = {Proceedings of the 35th Annual ACM Symposium on Applied Computing},
pages = {1026–1033},
numpages = {8},
location = {Brno, Czech Republic},
series = {SAC '20}
}

@article{survey,
  author       = {Ahmad Azab and Mahmoud Khasawneh and Saed Alrabaee and Kim-Kwang Raymond Choo and Maysa Sarsour},
  title        = {Network traffic classification: Techniques, datasets, and challenges},
  journal      = {Digital Communications and Networks},
  volume       = {10},
  number       = {3},
  pages        = {676--692},
  year         = {2024},
  issn         = {2352-8648},
  doi          = {10.1016/j.dcan.2022.09.009}
}

@misc{selenium,
  author       = {Selenium},
  title        = {Selenium - Web Browser Automation},
  year         = {2024},
  url          = {https://www.selenium.dev/},
}

@inproceedings{etbertcritique,
   title={SoK: Decoding the Enigma of Encrypted Network Traffic Classifiers},
   DOI={10.1109/sp61157.2025.00165},
   booktitle={2025 IEEE Symposium on Security and Privacy (SP)},
   publisher={IEEE},
   author={Wickramasinghe, Nimesha and Shaghaghi, Arash and Tsudik, Gene and Jha, Sanjay},
   year={2025},
   month=may, pages={1825–1843} }

@misc{softmax,
      title={A Baseline for Detecting Misclassified and Out-of-Distribution Examples in Neural Networks}, 
      author={Dan Hendrycks and Kevin Gimpel},
      year={2018},
      eprint={1610.02136},
      archivePrefix={arXiv},
      primaryClass={cs.NE},
}
\end{document}